\documentclass[twocolumn, showpacs, prl]{revtex4}

\usepackage{graphicx}

\usepackage{dcolumn}

\usepackage{bm}

\begin{document}

\title{Spin susceptibility of two-dimensional electrons in narrow AlAs quantum wells}

\date{\today}

\author{K. Vakili}

\author{Y. P. Shkolnikov}

\author{E. Tutuc}

\author{E. P. De Poortere}

\author{M. Shayegan}

\affiliation{Department of Electrical Engineering, Princeton
University, Princeton, NJ 08544}

\begin{abstract}

We report measurements of the spin susceptibility in dilute
two-dimensional electrons confined to a 45$\AA$ wide AlAs quantum
well.  The electrons in this well occupy an out-of-plane
conduction-band valley, rendering a system similar to
two-dimensional electrons in Si-MOSFETs but with only one valley
occupied. We observe an enhancement of the spin susceptibility
over the band value that increases as the density is decreased,
following closely the prediction of quantum Monte Carlo
calculations and continuing at finite values through the
metal-insulator transition.

\end{abstract}


\maketitle

As the density of a low-disorder two-dimensional electron system
(2DES) is lowered, the Coulomb energy dominates over the kinetic
(Fermi) energy and is expected to lead to a host of correlated
states at low temperatures.  According to quantum Monte Carlo
(QMC) calculations, for example, the spin susceptibility $\chi$
$\propto$ {\it g}*{\it m}* of the 2DES should rise and diverge
below a critical density at which the system attains a
ferromagnetic ground state ({\it g}* and {\it m}* are the
renormalized Land\'{e} g-factor and mass respectively)
\cite{attaccalite02}. Recent experiments \cite{okamoto99,
vitkalov01, shashkin01, pudalov02, tutuc02, zhu03, prus03,
tutuc03} in a number of different 2DESs have indeed revealed a
rise in {\it g}*{\it m}* with decreasing density, and there is even a report
of the divergence of {\it g}*{\it m}* at a density where the system exhibits a
metal-insulator transition \cite{shashkin01}. Almost none of the
measurements, however, agree quantitatively with the results of
QMC calculations. Moreover, in Si-MOSFETs it is reported that the
main contribution to the {\it g}*{\it m}* enhancement comes from
an increase in {\it m}* rather than {\it g}* \cite{shashkin02},
also contrary to the QMC prediction.

We report here measurements of {\it g}*{\it m}* and {\it m}* as a
function of density in high mobility 2DESs confined to very narrow
AlAs quantum wells (QWs).  In several respects, this is a nearly
ideal system to study the properties of a dilute 2DES: it is a
single-valley system with an isotropic in-plane Fermi contour and
a thin electron layer. We find that {\it g}*{\it m}* in this
system follows closely the prediction of recent QMC calculations
\cite{attaccalite02}.  The agreement continues through the
metal-insulator transition, and we observe no corresponding
ferromagnetic instability at that density. Our measurements of
{\it m}* exhibit a sample and cooldown dependence that betray a
possible inadequacy of the Dingle analysis in this system.

Bulk AlAs has an indirect band-gap with the conduction-band minima
at the six equivalent X-points of the Brillouin zone. The Fermi
surface consists of six, anisotropic, half-ellipsoids ({\it three}
full-ellipsoids or valleys), with transverse and longitudinal band
masses of {\it m}$_{t}$ $\simeq$ 0.2 and {\it m}$_{l}$ $\simeq$ 1
respectively. This is similar to Si except that Si has {\it six}
conduction-band valleys centered on points along the
$\Delta$-lines of the Brillouin zone. Electrons confined along the
growth direction in a narrow ($<$ 55 $\AA$) AlAs QW occupy the
{\it single} out-of-plane valley since its higher {\it m}* along
{\it z} gives a lower confinement energy than for the in-plane
valleys \cite{vankesteren89,yamada94,inplane}. This is again
similar to the 2DES in Si-MOSFETs fabricated on (001) Si
substrates, except that in the Si case {\it two} valleys are
occupied.

We performed measurements on four samples (A1, A2, A3, B1) from
two different wafers (A and B).  All were Si-modulation doped, 45
$\AA$-wide, AlAs QWs with Al$_{0.4}$Ga$_{0.6}$As barriers, grown
on GaAs (001) substrates. We patterned the samples in a Hall bar
configuration, and made ohmic contacts by depositing AuGeNi and
alloying in a reducing environment. Metallic front and back gates
were deposited to control the carrier density, {\it n}, which was
determined from the frequency of Shubnikov-de Haas (SdH)
oscillations and from the Hall resistance. Values of {\it n} in
our samples were in the range of 0.54 to 7.0 x 10$^{11}$
cm$^{-2}$, with a peak mobility of $\mu$ = 5.1 m$^{2}$/Vs in
sample A1; this is comparable to the highest mobilities reported
for Si-MOSFETs. Using the AlAs dielectric constant of 10, and the
transverse band effective mass {\it m}$_{b}$ = 0.21 reported from
cyclotron resonance measurements \cite{momose99}, our density
range corresponds to $2.8 \leq r_{s} \leq 9.6$, where {\it
r}$_{s}$ is the average inter-electron spacing measured in units
of the effective Bohr radius. We made measurements down to {\it T}
= 40 mK using low-frequency lock-in techniques. The samples were
mounted on a tilting stage to allow the angle,
$\theta$, between the normal to the sample and the magnetic field
to be varied {\it in situ}.

In Fig. 1, we exhibit a characteristic trace of longitudinal
resistivity at $\theta=0^{o}$. The high quality of the sample, as
compared to previous 2DESs in narrow AlAs QWs \cite{yamada94}, is
evident from the persistence of SdH oscillations to fields as low
as 0.5 T. The oscillations at the lowest fields correspond to
odd-integer Landau level (LL) filling factors ($\nu$), as
determined by the ratio of the Zeeman and cyclotron splittings
which, in turn, is fixed by the product {\it g}*{\it m}*. A simple
analysis reveals that the $\rho$$_{xx}$ minima at odd-integer
$\nu$ will be strong at the lowest fields when 4{\it j}+1 $<$ {\it
g}*{\it m}* $<$ 4{\it j}+3, for {\it j} a nonnegative integer;
this is the case for our 2DES as we will discuss below.

\begin{figure}
    \centering
    \includegraphics[scale=0.42]{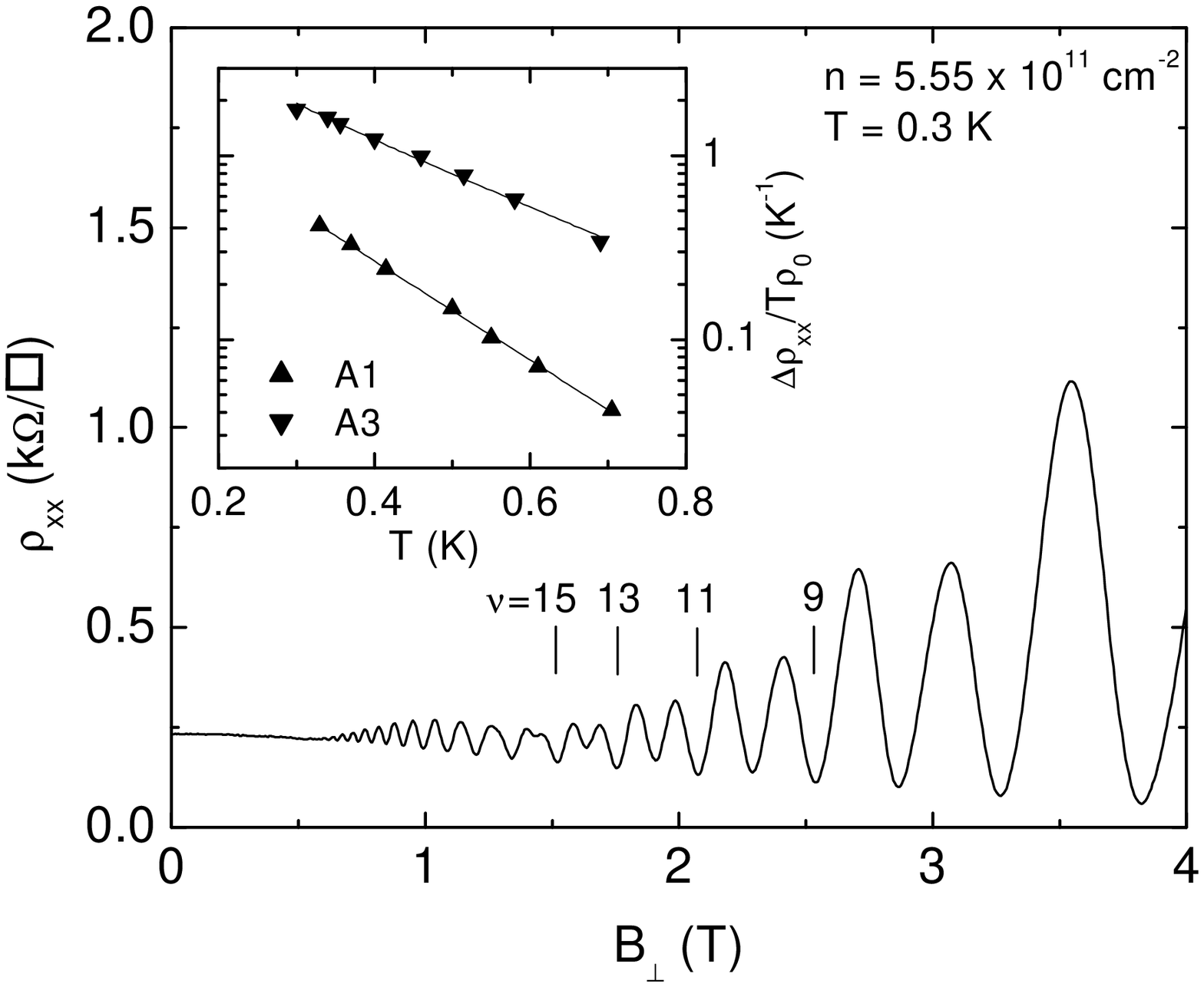}
    \caption{Longitudinal resistivity ($\rho$$_{xx}$) for sample A1 at {\it n} = 5.55x10$^{11}$ cm$^{-2}$.  Inset: Temperature dependence of the amplitude ($\Delta$$\rho$$_{xx}$) of SdH oscillations, normalized to the zero-field resistivity ($\rho$$_{0}$),
    at $\nu$ = 7 ({\it B} = 1.0 T) for samples A1 (up triangles, {\it n} = 1.82x10$^{11}$ cm$^{-2}$, $\mu$ = 4.7 m$^{2}$/Vs) and A3 (down triangles, {\it n} = 1.89x10$^{11}$ cm$^{-2}$, $\mu$ = 3.3 m$^{2}$/
    Vs).  The Dingle fits give effective masses of
    0.45 and 0.30 respectively.}
\end{figure}

For a quantitative determination of {\it g}*{\it m}*, we measured
magnetoresistance in fields applied parallel to the 2DES plane
($\theta$ = 90$^{o}$) to directly observe the total spin
polarization field at low densities (Fig. 2), and in tilted fields
(0$^{o}$$<$$\theta$$<$90$^{o}$) to determine {\it g}*{\it m}* at
low fields and high densities (Fig. 3). An example of our parallel
field measurements is displayed in Fig. 2. The traces are similar
to results from previous experiments \cite{vitkalov01, shashkin01,
okamoto99, tutuc02, zhu03}: a metallic {\it T}-dependence is
observed at zero magnetic field, followed by a field-induced
transition to an insulating behavior at high fields.  There is
also a resistance kink whose high field onset, marked by an arrow
({\it B}$_{P}$) in Fig. 2, signals the total spin polarization of
the carriers \cite{okamoto99, tutuc02, dolgopolov00}. In the inset
of Fig. 2, we have displayed {\it B}$_{P}$ versus {\it n} from our
parallel-field measurements. By monitoring {\it B}$_{P}$ as {\it
n} is changed, we can extract {\it g}*{\it m}* as a function of
{\it n} from the relation {\it B}$_{P}$ = (2{\it h}/{\it e}){\it
n}/{\it g}*{\it m}* (ignoring the possibility of nonlinear spin
polarization, which we will address shortly); these {\it g}*{\it
m}* are plotted in Fig. 4(a) (open symbols). The results are
normalized by the band values, {\it g}$_{b}$=2 and {\it
m}$_{b}$=0.21 \cite{vankesteren89, momose99}.

\begin{figure}
    \centering
    \includegraphics[scale=0.40]{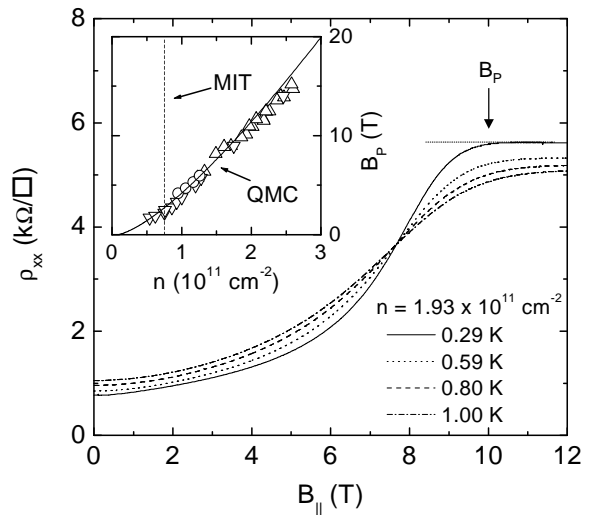}
    \caption{Parallel-field magnetoresistance at a fixed density for different
    temperatures. A field-induced transition from metallic to insulating behavior occurs at 7.7 T, and
    the 2DES becomes fully spin polarized at {\it B}$_{P}$ = 10.1 T (vertical arrow). Inset:
    Measured spin polarization fields ({\it B}$_{P}$) vs. density for samples A1 (up triangles), A3 (down triangles), and B1 (circles).
    The solid curve represents the quantum Monte Carlo (QMC) prediction for {\it B}$_{P}$ and the dotted line the zero-field metal-insulator transition (MIT) observed in our samples.}
\end{figure}

To augment our determination of {\it g}*{\it m}* at higher
densities, we next performed measurements in tilted fields. The
quantization of the 2DES energy into LLs depends on the component
of the magnetic field perpendicular to the plane of the 2DES,
while the Zeeman energy depends on the total field. By tilting the
sample relative to the applied field, we can change the ratio of
Zeeman and cyclotron splittings and therefore the strengths of
alternating SdH minima \cite{fang68}. The minima are weakest when
a set of spin-down LLs cross a set of spin-up LLs. This
corresponds to the condition, 1/cos($\theta$$_{c}$) = 2{\it
i}/{\it g}*{\it m}*, where $\theta$$_{c}$ is the angle at which
the levels cross and {\it i} is the difference in LL index between
the crossing levels. In Fig. 3 we have plotted, at a fixed
density, the strength of SdH minima, $\Delta$$\rho$$_{xx}$, for
several $\nu$ as a function of 1/cos($\theta$);
$\Delta$$\rho$$_{xx}$ is defined as the average difference between
the resistance at a SdH minimum and at the maxima on either side
of it. If the maxima become indiscernible, such as when a minimum
itself becomes a maximum, we simply take $\Delta$$\rho$$_{xx}$ =
0.  The LL crossings are indicated by the vertical dashed lines in
Fig. 3, with the even-$\nu$ minima experiencing a crossing at
$\theta$$_{c}$ = 57.7$^{o}$ and the odd-$\nu$ minima at
$\theta$$_{c}$ = 74.7$^{o}$. The lowest $\nu$ at which LL
crossings are observed is $\nu$=2, giving {\it i} = 1 for the
first observed crossing angle. The plot of 1/cos($\theta$$_{c}$)
versus {\it i}, shown in Fig. 3 inset, confirms the above LL
crossing condition. Using such analysis, we have extracted {\it
g}*{\it m}*/{\it g}$_{b}${\it m}$_{b}$ versus {\it n} and plotted
them in Fig. 4(a) (closed symbols). We note that the values of
{\it g}*{\it m}* derived from parallel and tilted field
measurements are consistent with each other and with the condition
discussed above for the odd filling SdH minima being strong at low
fields.

\begin{figure}
    \centering
    \includegraphics[scale=0.39]{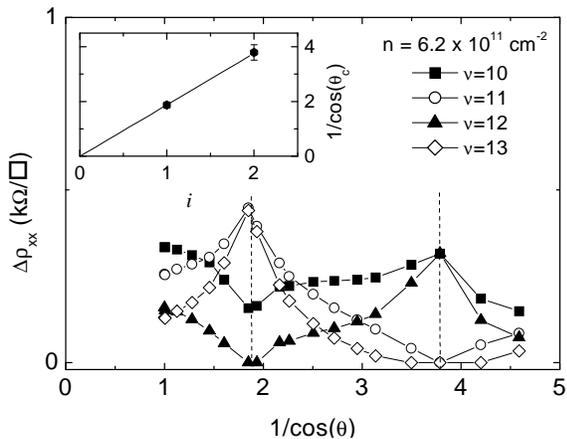}
    \caption{Strength of SdH minima as a function of tilt angle for several different filling factors.
    LL crossings are indicated by dashed lines, with the even-fillings experiencing a crossing at $\theta$$_{c}$ = 57.7$^{o}$
    and the odd at 74.7$^{o}$. Inset: Dependence of the LL crossing angle, $\theta_{c}$, on crossing index,
    {\it i}.}
\end{figure}

We now address the linearity of the spin polarization with field.
In their study of dilute GaAs 2DESs, Zhu {\it et al.} \cite{zhu03}
and Tutuc {\it et al.} \cite{tutuc03} report a significantly
nonlinear spin polarization as a function of magnetic field.
Measurements by Tutuc {\it et al.} \cite{tutuc03} demonstrate that
the finite thickness of the electron layer causes an enhancement
of the {\it single-particle} {\it m}* in parallel field. This
enhancement is significant when the magnetic length becomes
comparable to or less than the 2DES thickness and accounts for the
observed spin polarization nonlinearity in GaAs 2DESs in Ref.
\cite{tutuc03}. In contrast, we do not expect significant
finite-layer thickness effects for a 2DES in a narrow AlAs QW
given its very small width. Our results indeed reveal no
measurable nonlinear spin polarization with magnetic field. First,
the LL crossings occur at the same $\theta$$_{c}$ for all $\nu$ of
a given parity (Fig. 3) with only slight deviations at the very
lowest $\nu$. Second, the plots of 1/cos($\theta$$_{c}$) versus
{\it i} for various densities are all fit well by straight lines
that pass through the origin (Fig. 3 inset); this would not be the
case if the rate of polarization changed significantly in field.
Third, there is good agreement in our values of {\it g}*{\it m}*
derived from tilted-field LL crossings and parallel-field {\it
B}$_{P}$ measurements, although the former measurement is
performed at small, {\it partial} spin polarizations and low parallel
fields while the latter is done at {\it full} polarization and
high parallel fields.

Having established the linearity of spin polarization in this
system, we compare our {\it B}$_{P}$ and {\it g}*{\it m*} with the
prediction of a QMC calculation for an interacting 2DES which has
zero thickness and is disorder free \cite{attaccalite02}. That
prediction is shown in Fig. 2 inset and in Fig. 4(a)
\cite{manybody}.  The agreement of the measured and predicted
values is surprisingly good, considering that the calculation
ignores disorder.  While we cannot rule out that such good
agreement is fortuitous, we emphasize that there are essentially
no adjustable parameters in this comparison.  To convert the 2D
density into {\it r}$_{s}$, we used a band mass {\it m}$_{b}$ =
{\it m}$_{t}$ = 0.21; although there is some uncertainty regarding
the precise value of the band mass in AlAs, the 0.21 value is from
cyclotron resonance measurements \cite{momose99} and likely
represents the most accurate value currently available. We add
that we get good agreement between the QMC results and the
measured {\it g}*{\it m}* for 0.20 $\leq$ {\it m}$_{t}$ $\leq$
0.23. This range is also consistent with a recent determination of
{\it m}$_{t}$ from ballistic transport measurements in AlAs 2DESs
\cite{gunawan04}. On the other hand, SdH measurements of the {\it
m}*, including our own, tend to give slightly larger values. As we
discuss below, these latter measurements are problematic.

In our samples, we observe a zero-field metal-insulator transition
(MIT) at {\it n}=0.72 x 10$^{11}$ cm$^{-2}$ ({\it r}$_{s}$=8.3),
indicated by the vertical lines in the inset of Fig. 2 and
in Fig. 4. This is close to the value reported for wide AlAs QWs
with comparable mobilities but with the in-plane valley occupied
\cite{papadakis98}, and Si-MOSFETs with two out-of-plane valleys
occupied \cite{abrahams01}. More importantly, there is no apparent
divergence of {\it g}*{\it m}* at the MIT in our samples, and
instead the measured {\it g}*{\it m}* coincide with the QMC curve
continuously through the transition. The QMC calculation places
the ferromagnetic transition in our system at {\it n} $\sim$ 1 x
10$^{10}$ cm$^{-2}$, well below the range accessible in our
measurements.

\begin{figure}
    \centering
    \includegraphics[scale=0.52]{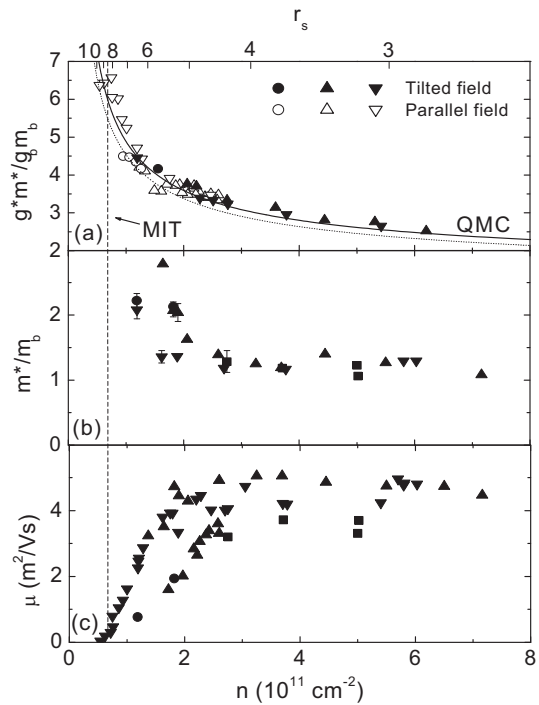}
    \caption{Shown as a function of density are: (a) spin susceptibility, (b) effective mass, and (c) mobility.  In
    all plots, up triangles correspond to sample A1, squares to A2, down triangles to A3, and circles to B1.
    The QMC prediction of {\it g}*{\it m}* at zero (dotted) and full (solid) polarization is included in (a), and the vertical dashed line indicates the density
    corresponding to the zero-field MIT.}
\end{figure}

We also performed independent measurements of {\it m}* as a
function of {\it n} [Fig. 4(b)] by monitoring the {\it
T}-dependence of the amplitude of low field SdH oscillations and
fitting the results with the Dingle formula (Fig. 1 inset).  Both
the mass and the Dingle temperature were allowed to vary as
fitting parameters. The masses that we include in Fig. 4(b) at
each {\it n} were obtained from data taken in the temperature
range 0.3 to 0.8 K and fields in the range of 0.50 to 1.5 T. We
have included two-sided error bars in Fig. 4(b) to reflect the
spread of {\it m}* in this field range. Our measured {\it m}* at
high {\it n} are comparable to those reported previously in
narrow-well AlAs \cite{yamada94, momose99}. At low densities,
there appears to be some enhancement of {\it m}* that is similar
in magnitude to that reported in Si-MOSFETs \cite{shashkin02,
pudalov02}, however there is a significant spread in the {\it m}*
values that depends on both sample and cooldown.  As examples, in
the inset of Fig. 1, we show Dingle plots for two different
samples at nearly the same field and density that give markedly
different values for {\it m}*.  It is noteworthy that the
discrepancies become large at approximately the same density where
the 2DES mobility begins to drop rapidly [Fig. 4(c)], though there
is no direct correlation between the mobilities and {\it m}*
values. The Dingle analysis may be producing anomalous results in
this system, particularly as the density is reduced, for several
reasons. At low densities, the zero-field resistance becomes
increasingly {\it T}-dependent, complicating the application of
the Dingle analysis which presupposes no such dependence.  We have
tried the analysis employed in Ref. \cite{pudalov02} whereby the
Dingle temperature is assumed to vary with {\it T} at the same
rate as the zero-field resistance; this fails to appreciably
reconcile the discrepancies between the {\it m}* values and simply
reduces the overall enhancement at low {\it n}. The discrepancies
could also be due to inhomogeneities in the 2DES density, which
can vary between samples and cooldowns and which would become
increasingly significant as the overall density is reduced.  In
any case, given that the discrepancies observed in {\it m}* are
absent in {\it g}*{\it m}*, it appears that the values deduced
from the Dingle analysis of the SdH oscillations do not reflect
the true {\it m}* at low densities in our samples.

We emphasize that the values of {\it g}*{\it m}* derived from
parallel and tilted field measurements are consistent between
samples and cooldowns. This is despite differences in mobility,
implying that the spin susceptibility is not very sensitive to the
effects of disorder. The continued agreement of our measurements
with the QMC calculation at low densities, where screening of the
disorder potential wanes and the role of disorder become more
important [Fig. 4(c)], is consistent with this observation.

We have argued that narrow AlAs QWs are ideal for measuring the
role of interaction on the spin susceptibility because of their
single-valley occupation, thin electron layer, and isotropic
in-plane Fermi contour.  These same properties may be responsible
for the differences between our results and those in other 2DESs.
Measurements of {\it g}*{\it m}* in Si-MOSFETs generally find less
enhancement over the band value than narrow AlAs QWs for most of
the available density range. Initially, this is somewhat
surprising considering that (001) Si-MOSFETs have a double-valley
occupancy which should produce a more dilute 2DES. However, recent
measurements in strain-tunable wide AlAs QWs find the same valley
occupation dependence of {\it g}*{\it m}* within a single material
system, namely less enhancement in the two-valley case
\cite{shkolnikov04}.

Another system of interest is the dilute GaAs 2DES where the
electrons occupy a single, isotropic valley. In Ref.
\cite{tutuc03}, where {\it g}*{\it m}* was determined via {\it
B}$_{P}$ measurements, it was shown that the differences between
the measured {\it g}*{\it m}* and the QMC results could be
essentially reconciled by taking into account the significant
Fermi surface distortion that occurs in a thick electron layer
system in the presence of a large parallel field.  Zhu {\it et
al.} \cite{zhu03}, however, have reported that even in the limit
of zero parallel field, the measured {\it g}*{\it m}*  remains
below the QMC results. While we do not have an explanation for
this observation, we speculate that the weaker enhancement may stem from  an effective reduction of the interaction strength at zero
field caused by the very large electron layer thickness in the
extremely dilute GaAs 2DESs of Ref.
\cite{zhu03}, rendering a system with a smaller effective {\it
r}$_{s}$ than the one calculated assuming zero thickness. A full,
quantitative understanding of the differences between the various
material systems is still lacking.

We thank the NSF for support, and R.\ Winkler for discussions.
Part of the work was performed at the Florida NHMFL, also
supported by the NSF; we thank E. Palm, T. Murphy, and G. Jones
for technical assistance.

\break


\begin{references}
\bibitem{inplane}
The in-plane valleys are occupied when the QW width is $>$ 55
$\AA$, so that the strain caused by the lattice mismatch between
AlAs and the GaAs substrate dominates the valley occupation
energies (see Refs. \cite{vankesteren89, yamada94}).
\end{references}
\end{document}